\documentclass[prb,twocolumn,showpacs,preprintnumbers,amsmath,amssymb]{revtex4}
\usepackage{setspace}
\usepackage{subfigure}
\usepackage{graphicx, epsfig}
\usepackage{dcolumn}
\def\be{\begin{equation}}
\def\ee{\end{equation}}
\def\bea{\begin{eqnarray}}
\def\eea{\end{eqnarray}}
\begin{document}

\title{First principles molecular dynamics study of amorphous Si$_{1-x}$Ge$_{x}$:H alloys}

\author{T. A. Abtew}
\email{abtew@phy.ohiou.edu}

\author{D. A. Drabold}
\email{drabold@ohio.edu}

\affiliation{Department of Physics and Astronomy, Ohio University, 
Athens, OH 45701}


\date{\today}
\begin{abstract}
We study the structural, dynamical and electronic properties of amorphous Si$_{1-x}$Ge$_{x}$:H alloys using first principles local basis molecular dynamics simulation. The network topology and defects of the amorphous network have been analyzed. Structural changes and an increase in number of defects have been found as the Ge atomic percentage increases from $x$=0.1 to $x$=0.5. The electronic properties show a narrower band-gap and increased mid-gap and band-tail defect states in the gap of the spectrum as Ge composition increases. Investigation on the band tails of the density of states show an exponential (Urbach) behavior. The mobility gap is predicted as a function of Ge concentration.

\end{abstract}
 
\maketitle

\section{Introduction}
\label{secI}

Hydrogenated amorphous Si-Ge alloys are important materials for uncooled microbolometer applications for night ``thermal" vision and IR sensing.\cite{bolometer} The materials are also of basic interest as a mild form of alloy disorder (here ``mild" refers to the chemical similarity of the two species) juxtaposed with topological disorder. Such alloys have been discussed for photovoltatic applications\cite{photoapp}, and features of the electronic structure in photoemission measurements\cite{aljishi}. While empirical models of amorphous phases have appeared\cite {ishimaru,mousseau},  no first principles simulations have been reported. Our aim is to offer small but reliable models of these important materials, compare them to experiment, and to make specific predictions of the structural origins of defect states appearing near the Fermi level.

The paper is organized as follows. In Sec.~\ref{secII}, we discuss the approximations and parameters used in the {\it ab initio} code employed here and describe procedures for generating a-Si$_{1-x}$Ge$_x$:H alloy models.  In Sec.~\ref{secIII} we describe the atomic structural properties by studying partial pair correlations, coordination and bond angle distribution of the network. The electronic properties of localized mid-gap and band-tail states are presented in Sec.~\ref{secIV}.  The dynamical properties and vibrational density of states are given in Sec.~\ref{secV}. Finally, we present conclusions.

\section{Methodology}
\label{secII}
\subsection{Total energy and force code}

The density functional calculations in the present work were performed within the generalized gradient approximation\cite{perdew} (GGA) using the first principles code SIESTA \cite{ordejon,sanchez,soler}. Calculations in this paper employed single $\zeta$ basis with polarization orbitals (SZP) for Si and Ge and a double $\zeta$ polarized basis (DZP) for hydrogen\cite{attafynn}. The structures were relaxed using conjugate gradient (CG) coordinate optimization until the forces on each atom were less than 0.02 eV/\AA. We used a plane wave cutoff of 100 Ry for the grid (used for computing multi-center matrix elements) with $10^{-4}$ for the tolerance of the density matrix  in self consistency steps. We solved the self-consistent Kohn-Sham equations by direct diagonalization of the Hamiltonian and a conventional mixing scheme. The $\Gamma (\vec{k}=0)$ point was used to sample the Brillouin zone in all calculations.

\subsection{Model Formation}

Realistic models of a-Si have been obtained from the WWW algorithm\cite{www}. To model hydrogenated structures, we developed a 223 atom a-Si:H model by removing two Si atoms and adding nine H atoms (of the 216-atom a-Si) to terminate all the dangling bonds except one (to enable the observation of one dangling bond defect). Each of the H atoms were placed about 1.5~\AA~from the corresponding 3-fold Si atom. After relaxation, we replaced some of the Si atoms  by Ge atoms at random, and then relaxed the alloy using conjugate gradient minimization to generate a-Si$_{1-x}$Ge$_{x}$:H alloys, with the Ge fraction $x$ being 0.1, 0.2, 0.3, 0.4 and 0.5. Note that such clustered H agrees with the proton NMR second moment data in a-Si:H\cite{pafdad93}.

\section{Atomic Structural Properties}
\label{secIII}
\subsection{Pair correlation functions}

The topology of models may be analyzed by partial pair correlation functions $g_{\alpha \beta}(r)$ of atomic species $\alpha$ and $\beta$. The partial pair correlation $g_{\alpha \beta}(r)$ can be written as 
  		   \begin{equation}
 			   g_{\alpha \beta}(r)= \frac{1}{4\pi r^2 \rho N c_{\alpha} c_{\beta}}\sum_{i \ne j}\delta({r}-{r}_{ij})
  		  \end{equation}
where $N$ is the total number of particles in the system; $\rho = \frac{N}{V}$ is the number density , $c_{\alpha} = \frac{N_{\alpha}}{N}$ and $c_{\beta} = \frac{N_{\beta}}{N}$.  
We have used a Gaussian approximation for the delta function with broadening $\sigma = 0.1~$\AA.  

\begin{figure}[h]
\begin{center}
\includegraphics[angle=0,width=0.5\textwidth]{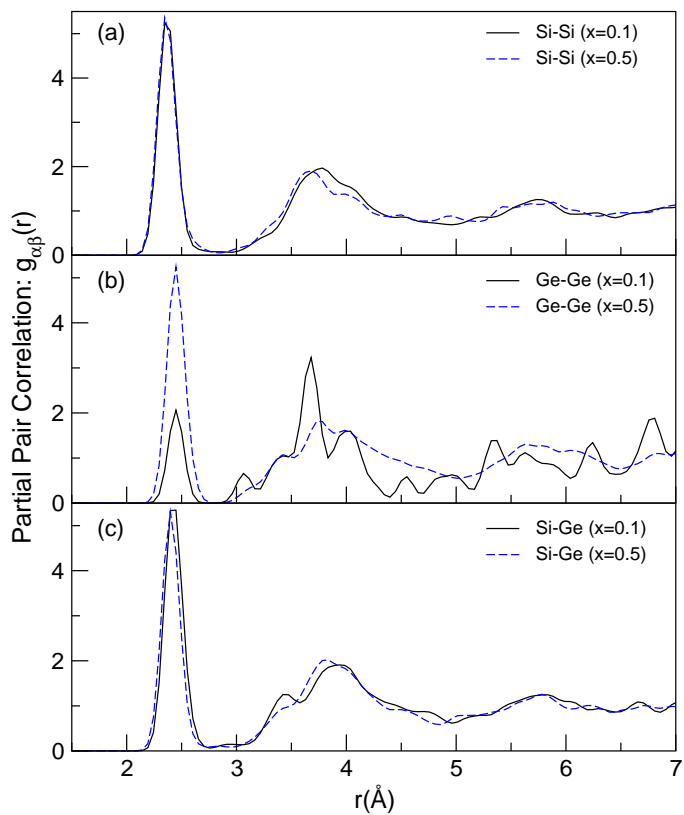}
\caption{\label{fig1} (Color Online) Partial pair distribution function $g_{\alpha \beta}$ of a-Si$_{1-x}$Ge$_{x}$:H alloys for two ($x$=0.1 and $x$=0.5) Ge atomic percentage of (a) Si\textendash Si, (b) Ge\textendash Ge, and (c) Si\textendash Ge.   } 
\end{center}
\end{figure}

We have analyzed  the five alloy compositions,  and a clear first nearest neighbor peak with subsequent deep minimum is evident. In Fig.~\ref{fig1} we plot the partial pair correlation for Si\textendash Si, Ge\textendash Ge, and Si\textendash Ge in the a-Si$_{1-x}$Ge$_{x}$:H alloy for $x$=0.1 and $x$=0.5.  As the Ge content increases from $x$=0.1 to $x$=0.5, the peak is fixed at $r_o\sim$2.50 \AA. The partial correlation for Si\textendash Ge has a peak at $r_o\sim$2.4 \AA. For the same Ge atomic percentage, the partial pair correlation of H\textendash H, Si\textendash H, and Ge\textendash H show a first peak $r_o$ of $\sim$1.45~\AA, $\sim$1.5~\AA~and $\sim$1.6~\AA~respectively. These results reproduce a trend observed in experiment \cite{aldrich}.  

\subsection{Geometry of Bonding}

We obtained partial $n_{\text{Ge}}$, $n_{\text{Si}}$, $n_{\text{H}}$, and average coordination numbers, $n$, based on the nearest neighbor distance determined in the preceding section, first neighbor coordination numbers $n_{\text{SiSi}}$, $n_{\text{SiGe}}$, $n_{\text{SiH}}$, $n_{\text{GeGe}}$, $n_{\text{GeSi}}$, and $n_{\text{GeH}}$ are obtained by integrating the pair correlation function $4\pi r^2\rho g_{\alpha\beta}$(r). The results are shown in Table~\ref{Table1}. The total coordination numbers for Ge, Si are,  $n_{\text{Ge}}$ = $n_{\text{GeGe}}$ + $n_{\text{GeSi}}$ + $n_{\text{GeH}}$ and $n_{\text{Si}}$ = $n_{\text{SiSi}}$ + $n_{\text{SiGe}}$ + $n_{\text{SiH}}$ respectively. We have obtained 1.0 for total coordination of H in all cases. 
 
\begin{table}[!hbp]
\caption{\label{Table1} The value of $r_o$ in the first peak of the g(r) and the first neighbor coordination number $n_{\alpha \beta}$ in a-Si$_{1-x}$Ge$_{x}$:H alloys for five different Ge atomic compositions $x$=0.1\textendash 0.5. The integration ranges are from 0.0\textendash 2.8 \AA~for Si\textendash Si, Ge\textendash Ge, Si\textendash Ge, Ge\textendash Si, and 0.0\textendash 1.8 \AA~for Si\textendash H, Ge\textendash H, H\textendash Si, and H\textendash Ge.}

\begin{tabular*}{0.48\textwidth}{@{\extracolsep{\fill}}lcccccc}
\hline\hline
\\
\multicolumn{1}{c}{} &
\multicolumn{6}{c}{n$_{\alpha \beta}$ for first shell} \\ \cline{3-7} 
 \multicolumn{1}{l}{Bond} &
\multicolumn{1}{c}{} &
\multicolumn{1}{c}{}&
\multicolumn{1}{c}{}&
\multicolumn{1}{c}{}&
\multicolumn{1}{c}{}&
\multicolumn{1}{c}{} \\
    \multicolumn{1}{l}{type} &
    \multicolumn{1}{c}{$r_o$(\AA)} &
    \multicolumn{1}{c}{ \small{($x$=0.1)}} &
    \multicolumn{1}{c}{ \small{($x$=0.2)}} &
    \multicolumn{1}{c}{ \small{($x$=0.3)}} &
    \multicolumn{1}{c}{ \small{($x$=0.4)}} &
    \multicolumn{1}{c}{ \small{($x$=0.5)}} \\ \cline{1-7}
    \\
\small{Si\textendash Si}		& 2.37 	& 3.47	& 3.07  	& 2.70       & 2.37       & 2.06   \\
\small{Si\textendash Ge}		& 2.42  	& 0.50  	& 0.86	& 1.22       & 1.54       & 1.97   \\ 
\small{Si\textendash H}		& 1.53  	& 0.04 	& 0.05	& 0.05       & 0.05       & 0.06   \\
\\
\small{Ge\textendash Ge}		& 2.46  	& 0.16  	& 0.71	& 1.11       & 1.58       & 1.92   \\ 
\small{Ge\textendash Si}		& 2.42  	& 3.80  	& 3.22	& 2.80       & 2.34       & 2.04   \\
\small{Ge\textendash H}		& 1.60  	& 0.04  	& 0.02	& 0.03       & 0.02       & 0.03   \\
\\
\small{H\textendash Si}		& 1.53  	& 0.89  	& 0.89	& 0.78       & 0.78      & 0.67   \\
\small{H\textendash Ge}		& 1.60  	& 0.11	& 0.11	& 0.22       & 0.22      & 0.33   \\
\\
\hline \hline
\end{tabular*}
\end{table}

To investigate the effect of Ge composition on the structures, we analyzed and obtained all types of bonding and structures in the network for each Ge compositions considered. For $x$=0.1, about 96.81\% of Si and 95.83\% of Ge are fourfold, only 1.06\% of Si and 4.17\% of Ge are threefold coordinated, 2.13\% of Si are fivefold. No fivefold coordination is obtained for Ge. Where fourfold Si is concerned, Si$_4$ is a dominant structure which is followed by Si$_3$Ge. We observed a similar pattern in the Ge fourfold coordination that Ge bonded to Si$_4$ structure is highly dominant which is followed by Ge bonded with Si$_3$Ge. The detailed results are shown in Table~\ref{Table2}. 

 \begin{table}
\caption{\label{Table2} Average percentage $m_\alpha(l)$ (bold characters) of atoms of species Si and Ge, $l$ fold coordinated at a distance of 2.68 \AA~for both Si and Ge, and 1.5 \AA~for H in a-Si$_{1-x}$Ge$_{x}$:H alloy for Ge atomic composition $x$=0.1. We also give the identity and the number of Ge and Si neighbors for each value of $m_\alpha(l)$.}
\begin{tabular*}{0.48\textwidth}{@{\extracolsep{\fill}}llllll}
\hline\hline
\\
    \multicolumn{1}{l}{Si } &
    \multicolumn{1}{l}{}   &
    \multicolumn{1}{l}{} &
    \multicolumn{1}{l}{}   &
    \multicolumn{1}{l}{\it l = 3}&
    \multicolumn{1}{l}{\bf 1.06} \\
    
    \multicolumn{1}{l}{ } &
    \multicolumn{1}{l}{}   &
    \multicolumn{1}{l}{} &
    \multicolumn{1}{l}{}   &
    \multicolumn{1}{l}{Si$_3$}&
    \multicolumn{1}{l}{0.53} \\
    
    \multicolumn{1}{l}{ } &
    \multicolumn{1}{l}{}   &
    \multicolumn{1}{l}{} &
    \multicolumn{1}{l}{}   &
    \multicolumn{1}{l}{Si$_2$Ge}&
    \multicolumn{1}{l}{0.53} \\
    
    \multicolumn{1}{l}{\it  l = 4} &
    \multicolumn{1}{r}{\bf 96.81}   &
    \multicolumn{1}{l}{\it l = 5} &
    \multicolumn{1}{l}{\bf 2.13}   &
    \multicolumn{1}{l}{}&
    \multicolumn{1}{l}{} \\
    
     \multicolumn{1}{l}{ Si$_4$} &
    \multicolumn{1}{r}{52.67}   &
    \multicolumn{1}{l}{Si$_5$} &
    \multicolumn{1}{l}{1.06}   &
    \multicolumn{1}{l}{}&
    \multicolumn{1}{l}{} \\      
    
     \multicolumn{1}{l}{ Si$_3$Ge} &
    \multicolumn{1}{r}{32.98}   &
    \multicolumn{1}{l}{Si$_4$Ge} &
    \multicolumn{1}{l}{1.06}   &
    \multicolumn{1}{l}{}&
    \multicolumn{1}{l}{} \\      
    
     \multicolumn{1}{l}{ Si$_2$Ge$_2$} &
    \multicolumn{1}{r}{6.91}   &
    \multicolumn{1}{l}{} &
    \multicolumn{1}{l}{}   &
    \multicolumn{1}{l}{}&
    \multicolumn{1}{l}{} \\    
    
     \multicolumn{1}{l}{ Si$_3$H} &
    \multicolumn{1}{r}{3.72}   &
    \multicolumn{1}{c}{} &
    \multicolumn{1}{c}{}   &
    \multicolumn{1}{c}{}&
    \multicolumn{1}{c}{} \\      
    
     \multicolumn{1}{l}{ Si$_2$GeH} &
    \multicolumn{1}{r}{0.53}   &
    \multicolumn{1}{c}{} &
    \multicolumn{1}{c}{}   &
    \multicolumn{1}{c}{}&
    \multicolumn{1}{c}{} \\    
    \\
    \hline
    \\
    \multicolumn{1}{l}{Ge} &
    \multicolumn{1}{c}{}   &
    \multicolumn{1}{l}{\it l = 3} &
    \multicolumn{1}{r}{\bf 4.17}   &
    \multicolumn{1}{l}{\it l = 4}&
    \multicolumn{1}{l}{\bf 95.83} \\
    
    \multicolumn{1}{l}{ } &
    \multicolumn{1}{c}{}   &
    \multicolumn{1}{l}{Si$_3$} &
    \multicolumn{1}{r}{4.17}   &
    \multicolumn{1}{l}{Si$_4$}&
    \multicolumn{1}{r}{79.17} \\
    
     \multicolumn{1}{l}{ } &
    \multicolumn{1}{c}{}   &
    \multicolumn{1}{c}{} &
    \multicolumn{1}{c}{}   &
    \multicolumn{1}{l}{Si$_3$Ge}&
    \multicolumn{1}{r}{12.50} \\
    
    \multicolumn{1}{l}{ } &
    \multicolumn{1}{c}{}   &
    \multicolumn{1}{c}{} &
    \multicolumn{1}{c}{}   &
    \multicolumn{1}{l}{Si$_2$GeH}&
    \multicolumn{1}{r}{4.17} \\          
    
    \\
\hline \hline
\end{tabular*}
\end{table}

 In the case of $x$=0.5, we observed $\sim$10.47\% fivefold bonds for Si. About 87.21\% of Si and 99.04\% of Ge are fourfold, only 2.33\% of Si and 0.96\% of Ge are threefold coordinated. The dominant structure in this case is Si atom bonded to Si$_2$Ge$_2$ which is followed by Si bonded with Si$_3$Ge. Similar pattern is observed in the Ge fourfold coordination that Ge bonded to Si$_2$Ge$_2$ is the primary structure which is followed by Ge bonded with Si$_3$Ge. The detailed results are shown in Table~\ref{Table3}. Comparing the bonding statistics with the initial configuration, where we have randomly substituted Si atoms by Ge, upon relaxation we have observed: a decrease in the number of four folded atoms, an increase in the number of strained bond (2.5~\AA$<r<$2.7~\AA), $\sim 9.5\%$ for the case of $x$=0.5, and decrease in the normal bonds ($r<$ 2.5~\AA). 
   
\begin{table}
\caption{\label{Table3} Average percentage $m_\alpha(l)$ (bold characters) of atoms of species Si and Ge, $l$ fold coordinated at a distance of 2.65 \AA~for both Si and Ge, and 1.5 \AA~for H in a-Si$_{1-x}$Ge$_{x}$:H alloy for Ge atomic composition $x$=0.5. We also give the identity and the number of Ge and Si neighbors for each value of $m_\alpha(l)$.}
\begin{tabular*}{0.48\textwidth}{@{\extracolsep{\fill}}llllll}
\hline\hline
\\
    \multicolumn{1}{l}{Si } &
    \multicolumn{1}{l}{}   &
    \multicolumn{1}{l}{} &
    \multicolumn{1}{r}{}   &
    \multicolumn{1}{l}{\it l = 3}&
    \multicolumn{1}{r}{\bf 2.32} \\
    
    \multicolumn{1}{l}{ } &
    \multicolumn{1}{l}{}   &
    \multicolumn{1}{l}{} &
    \multicolumn{1}{r}{}   &
    \multicolumn{1}{l}{Si$_2$Ge}&
    \multicolumn{1}{r}{1.16} \\

     \multicolumn{1}{l}{ } &
    \multicolumn{1}{l}{}   &
    \multicolumn{1}{l}{} &
    \multicolumn{1}{r}{}   &
    \multicolumn{1}{l}{Si$_3$}&
    \multicolumn{1}{r}{1.16} \\    
    \\
    \multicolumn{1}{l}{\it  l = 4} &
    \multicolumn{1}{r}{\bf 87.21}   &
    \multicolumn{1}{l}{\it l = 5} &
    \multicolumn{1}{l}{\bf 10.47}   &
    \multicolumn{1}{l}{}&
    \multicolumn{1}{l}{} \\
    
     \multicolumn{1}{l}{ Si$_2$Ge$_2$} &
    \multicolumn{1}{r}{51.16}   &
    \multicolumn{1}{l}{Si$_4$Ge$_1$} &
    \multicolumn{1}{r}{3.49}   &
    \multicolumn{1}{l}{}&
    \multicolumn{1}{l}{} \\      
    
     \multicolumn{1}{l}{Si$_3$Ge$_1$} &
    \multicolumn{1}{r}{19.77}   &
    \multicolumn{1}{l}{Si$_3$Ge$_2$} &
    \multicolumn{1}{r}{5.81}   &
    \multicolumn{1}{l}{}&
    \multicolumn{1}{l}{} \\      
    
    \multicolumn{1}{l}{ Ge$_4$} &
    \multicolumn{1}{r}{8.14}   &
    \multicolumn{1}{l}{Si$_4$H} &
    \multicolumn{1}{r}{1.16}   &
    \multicolumn{1}{l}{}&
    \multicolumn{1}{l}{} \\    
    
     \multicolumn{1}{l}{ Si$_4$} &
    \multicolumn{1}{r}{5.81}   &
    \multicolumn{1}{l}{} &
    \multicolumn{1}{l}{}   &
    \multicolumn{1}{l}{}&
    \multicolumn{1}{l}{} \\    
    
     \multicolumn{1}{l}{ Si$_3$H} &
    \multicolumn{1}{r}{1.16}   &
    \multicolumn{1}{c}{} &
    \multicolumn{1}{c}{}   &
    \multicolumn{1}{c}{}&
    \multicolumn{1}{c}{} \\      
    
     \multicolumn{1}{l}{ Si$_2$GeH} &
    \multicolumn{1}{r}{1.16}   &
    \multicolumn{1}{c}{} &
    \multicolumn{1}{c}{}   &
    \multicolumn{1}{c}{}&
    \multicolumn{1}{c}{} \\    
    \\
    \hline
    \\
    \multicolumn{1}{l}{Ge} &
    \multicolumn{1}{c}{}   &
    \multicolumn{1}{l}{} &
    \multicolumn{1}{r}{}   &
    \multicolumn{1}{l}{\it l = 3}&
    \multicolumn{1}{r}{\bf 0.96} \\
    
    \multicolumn{1}{l}{ } &
    \multicolumn{1}{c}{}   &
    \multicolumn{1}{l}{} &
    \multicolumn{1}{r}{}   &
    \multicolumn{1}{l}{Si$_2$Ge}&
    \multicolumn{1}{r}{0.96} \\
    \\
    \multicolumn{1}{l}{ \it l = 4} &
    \multicolumn{1}{c}{\bf 99.04}   &
    \multicolumn{1}{c}{} &
    \multicolumn{1}{c}{}   &
    \multicolumn{1}{l}{}&
    \multicolumn{1}{l}{} \\
    
    \multicolumn{1}{l}{Si$_2$Ge$_2$} &
    \multicolumn{1}{r}{40.38}   &
    \multicolumn{1}{c}{}  &
    \multicolumn{1}{c}{}  &
    \multicolumn{1}{l}{}   &
    \multicolumn{1}{l}{} \\
    
    \multicolumn{1}{l}{Si$_3$Ge} &
    \multicolumn{1}{r}{26.92}   &
    \multicolumn{1}{c}{}  &
    \multicolumn{1}{c}{}  &
    \multicolumn{1}{l}{}   &
    \multicolumn{1}{l}{} \\
    
    \multicolumn{1}{l}{SiGe$_3$} &
    \multicolumn{1}{r}{21.15}   &
    \multicolumn{1}{c}{}  &
    \multicolumn{1}{c}{}  &
    \multicolumn{1}{l}{}   &
    \multicolumn{1}{l}{} \\
    
    \multicolumn{1}{l}{Si$_4$} &
    \multicolumn{1}{r}{3.85}   &
    \multicolumn{1}{c}{}  &
    \multicolumn{1}{c}{}  &
    \multicolumn{1}{l}{}   &
    \multicolumn{1}{l}{} \\    
     
     \multicolumn{1}{l}{Ge$_4$} &
    \multicolumn{1}{r}{3.85}   &
    \multicolumn{1}{c}{}  &
    \multicolumn{1}{c}{}  &
    \multicolumn{1}{l}{}   &
    \multicolumn{1}{l}{} \\  
    
    \multicolumn{1}{l}{Ge$_2$SiH} &
    \multicolumn{1}{r}{1.92}   &
    \multicolumn{1}{c}{}  &
    \multicolumn{1}{c}{}  &
    \multicolumn{1}{l}{}   &
    \multicolumn{1}{l}{} \\

     \multicolumn{1}{l}{Ge$_3$H} &
    \multicolumn{1}{r}{0.96}   &
    \multicolumn{1}{c}{}  &
    \multicolumn{1}{c}{}  &
    \multicolumn{1}{l}{}   &
    \multicolumn{1}{l}{} \\             
    \\
\hline \hline
\end{tabular*}
\end{table}
        
\subsubsection{Angular Distribution}

\begin{figure}[!hbp]
\begin{center}
\includegraphics[angle=0,width=0.5\textwidth]{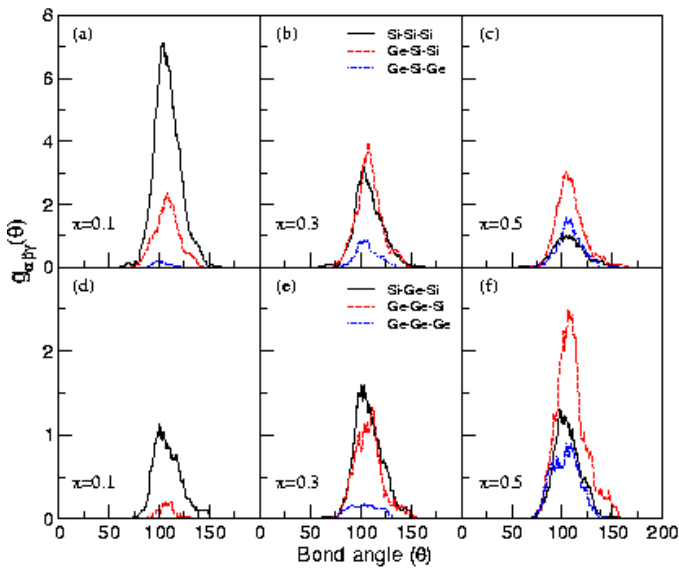}
\caption{\label{fig2} (Color Online)  The partial bond-angle distribution function as a function of bond angle $\theta$ in a-Si$_{1-x}$Ge$_x$:H for $x$=0.1 (left panel), for $x$=0.3 (middle panel), and for $x$=0.5 (right panel). (a), (b) and (c) are partial angular distribution for three possible angles centering Si and (d), (e), and (f) are partial angles taking Ge as a center.} 
\end{center}
\end{figure}

We have calculated the partial angular distribution for a-Si$_{1-x}$Ge$_x$:H with $x$=0.1, $x$=0.3, and $x$=0.5 Ge compositions and plotted in Fig.~\ref{fig2}(a)-(f). Though we chose only the three Ge compositions the results reported here are similar to the other two Ge compositions $x$=0.2 and $x$=0.4.
The partial pair correlation functions for $\alpha$\textendash Si\textendash$\gamma$ are plotted in the upper panel and the partial pair correlation functions for  $\alpha$\textendash Ge\textendash$\gamma$ are plotted in the lower panel. 
In each of the cases considered, we found total angular distribution peaks centered near the tetrahedral angle with $\theta$ in the range 103$^\circ$-110$^\circ$. 

Analysis of different Ge concentrations showed similar total angular distribution (the sum of all the partial angular distributions) because of the similarity in chemical properties and the proximity in different bond-lengths of Si\textendash Si, Si\textendash Ge, and Ge\textendash Ge. However, as mentioned above, the different compositions affects the type of bonds and structures formed in the network. There are also other partial angular distributions coming from H. 
The mean partial angular distribution of $\text{H}$\textendash Si\textendash$\text{Si}$ and $\text{H}$\textendash Ge\textendash$\text{Ge}$ are close to the tetrahedral angle 109.47$^\circ$ while the other two partials, $\text{H}$\textendash Ge\textendash$\text{Si}$ and $\text{H}$\textendash Si\textendash$\text{Ge}$, are off from a tetrahedral angle and ranges from 100.0$^\circ$\textendash 116.0$^\circ$. 
The probability density for cos($\theta$) is normal (Gaussian).  Normally distributed cosines of bond angles has been associated with exponential band tails in the electron density of states near the valence and conduction edges\cite{jianjun}. 

\section{ELECTRONIC STRUCTURE}
\label{secIV}
\subsection{Density of states}

Electronic structure has been described by the electronic density of states (EDOS), which was obtained by summing suitably broadened Gaussian centered at each Kohn-Sham eigenvalue.  In the five different Ge compositions we analyzed the EDOS. The results showed a narrow band-gap spectrum $\sim$1.6 eV for $x$=0.1, that becomes narrower as the Ge composition increases ($\sim$0.8 eV for $x$=0.5). This in turn introduces mid-gap and band-tail defect states in the spectrum which narrows the band-gap.

\begin{figure}[htbp]
\begin{center}
\includegraphics[angle=0,width=0.48\textwidth]{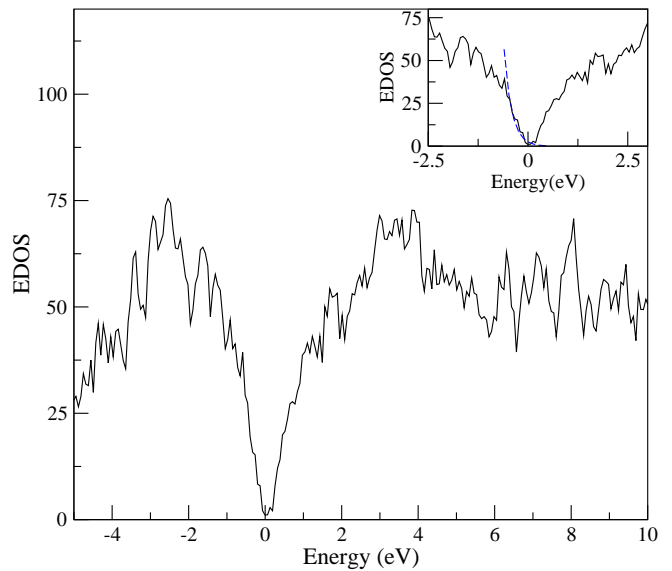}
\caption{\label{fig3}(Color Online) The electron density of states for a-Si$_{1-x}$Ge$_x$:H for $x$=0.4. The exponential fit in the inset for the valence band tail is 2.5$\times e^{-E/E_o}$ with $E_o = 192$ meV for $x$=0.4. The Fermi level is shifted to E=0.    } 
\end{center}
\end{figure}

In this section, we present the results of one of the alloys (for a-Si$_{1-x}$Ge$_x$:H with $x$=0.4) among the five different Ge atomic compositions. The electron density of states (EDOS) shown in Fig.~\ref{fig3} shows a narrow gap. The band tails of the spectrum which we take in the region (-0.7\textendash 0.0 eV ) fits exponential with $\sim e^{-E/E_o}$ with $E_o = 192$ meV as shown in the inset of Fig.~\ref{fig3}. We analyze these defect states in the spectrum with detail in the next sections.

\subsection{Localization: Inverse participation ratio}
\label{secVb}
In order to understand the electron localization we used the inverse participation ratio, IPR, 
 \begin{equation}
 \textrm{IPR} = \sum_{i=1}^{N}[q_i(E)]^2
 \end{equation}
where $q_i(E)$ is the Mulliken charge residing at an atomic site $i$ for an eigenstate with eigenvalue $E$ that satisfies $\sum_{i}^{N}[q_i(E)] = 1$ and $N$ is the total number of atoms in the cell. For an ideally localized state, only one atomic site contributes all the charge and so IPR=1. For a uniformly extended state, the Mulliken charge contribution per site is uniform and equals $1/N$ and so IPR=$1/N$. Thus, large IPR corresponds to localized states. 

\begin{figure}
\begin{center}
\includegraphics[angle=0,width=0.48\textwidth]{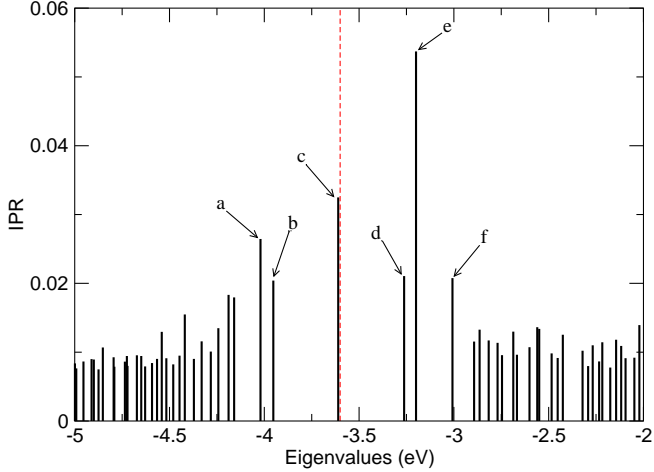}
\caption{\label{fig4}(Color Online)  Inverse participation ratio (IPR) in the a-Si$_{1-x}$Ge$_x$:H alloy for $x$=0.1 versus the energy eigenvalues. The dashed line is the Fermi level. } 
\end{center}
\end{figure}

\begin{figure}
\begin{center}
\includegraphics[angle=0,width=0.48\textwidth]{fig5}
\caption{\label{fig5} (Color Online) The contribution of atoms to the IPR (10\% and above) of a given state in a-Si$_{1-x}$Ge$_x$:H alloy for $x$=0.1. The labels from $a$-$f$ corresponds to different mid-gap and band-tail states of Fig.~\ref{fig4}.  } 
\end{center}
\end{figure}

\begin{figure}
\begin{center}
\includegraphics[angle=0,width=0.45\textwidth]{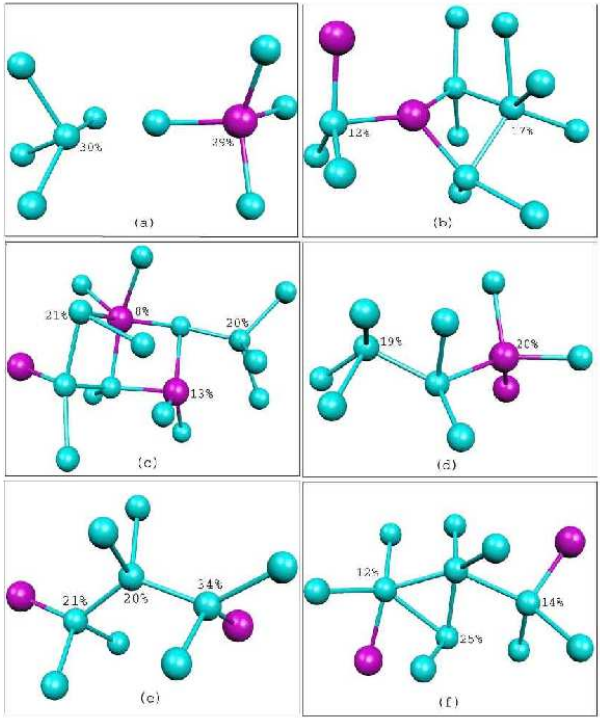}

\caption{\label{fig6}(Color Online) Structures which contributes 10\% and above to the inverse participation ratio that corresponds to the mid-gap and band-tail states shown in Fig.~\ref{fig4} from $a\textendash f$ in the a-Si$_{1-x}$Ge$_x$:H alloy for $x$=0.1. The color codes are blue for Si atoms and red for Ge atoms. } 
\end{center}
\end{figure}

\begin{figure}
\begin{center}
\includegraphics[angle=0,width=0.48\textwidth]{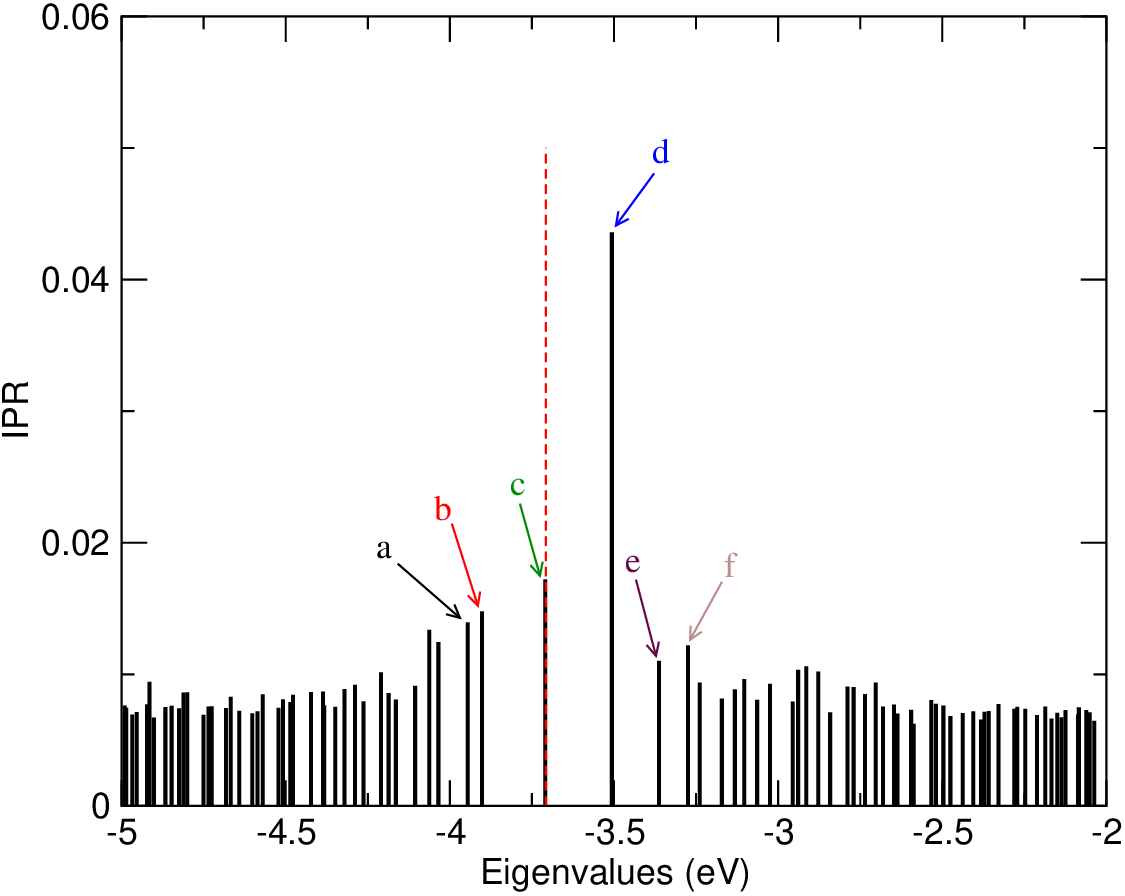}
\caption{\label{fig7}(Color Online) Inverse participation ratio (IPR) versus the energy eigenvalues in the a-Si$_{1-x}$Ge$_x$:H alloy for $x$=0.4. The dashed line is the Fermi level.   } 
\end{center}
\end{figure}

\begin{figure}
\begin{center}
\includegraphics[angle=0,width=0.48\textwidth]{fig8}
\caption{\label{fig8} (Color Online) The contribution of atoms to the IPR (5\% and above) of a given state in a-Si$_{1-x}$Ge$_x$:H alloy for $x$=0.4. The labels from $a\textendash f$ corresponds to different mid-gap and band-tail states of Fig.~\ref{fig7}.  } 
\end{center}
\end{figure}

\begin{figure}
\begin{center}
\includegraphics[angle=0,width=0.45\textwidth]{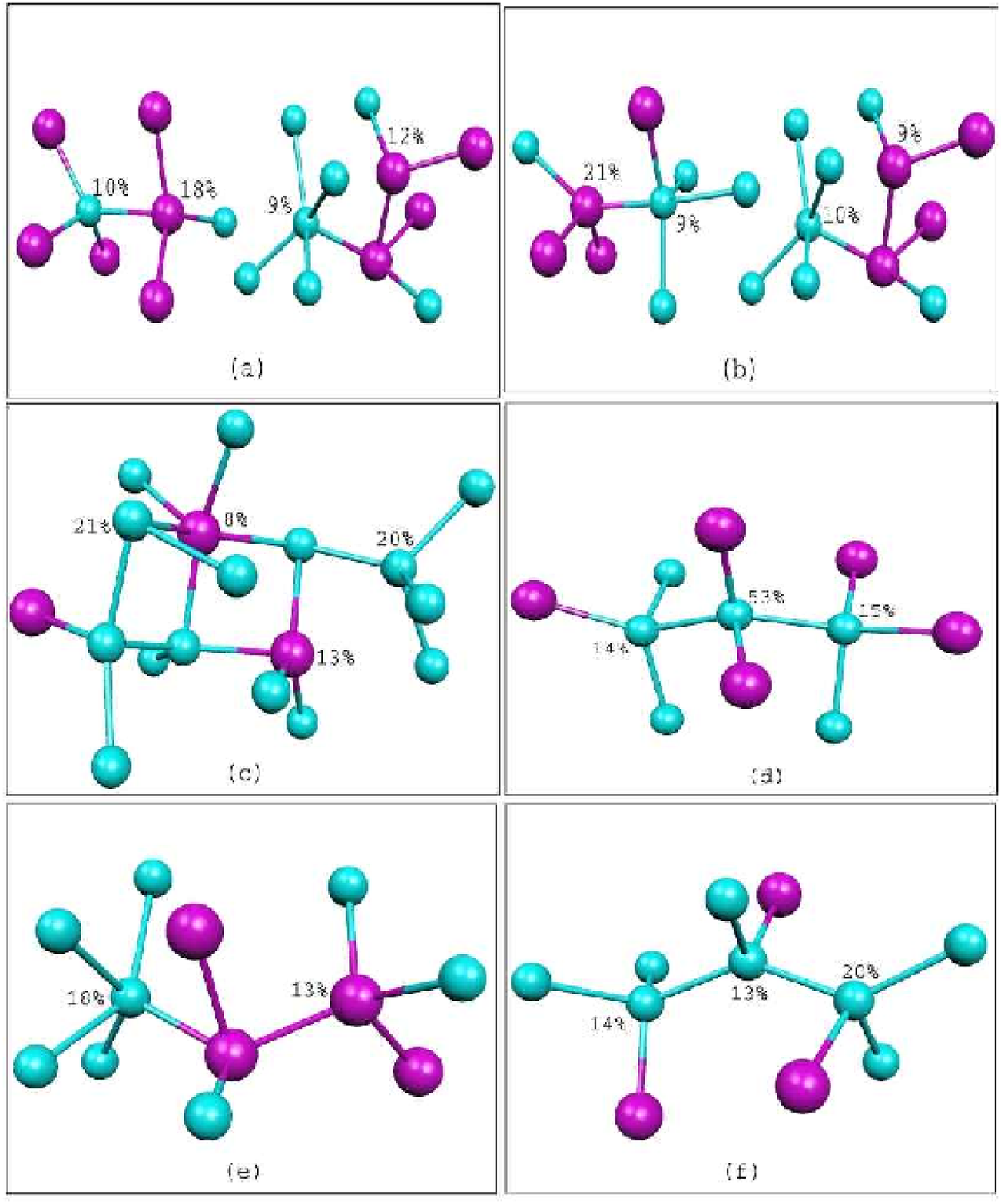}

\caption{\label{fig9} (Color Online) Structures corresponding to the mid-gap and band-tail states shown in Fig.~\ref{fig7} from $a\textendash f$ in the a-Si$_{1-x}$Ge$_x$:H alloy for $x$=0.4. The color codes are Si atoms (blue) and Ge atoms (red).} 
\end{center}
\end{figure}

To investigate how the electronic properties change in the gap, we have calculated the IPR of a-Si$_{1-x}$Ge$_x$:H alloy for two different Ge compositions, $x$=0.1 and $x$=0.4.  In each case, we calculated the IPR of the alloy in its relaxed ground state. We have also obtained contributions of each of the atoms in the alloy to the total IPR of a particular eigenstate. This allow us to see different types and geometries of structures which are responsible for generating a given state. Since we are interested in band-tail and mid-gap states, we narrowed our presentation here only to those eigenstates which are in the mid-gap or near the band-tails of the spectrum.

We plotted the inverse participation ratio and the contributions of each of the atoms to the IPR for $x$=0.1 are plotted in Fig.~\ref{fig4} and Fig.~\ref{fig5} respectively. For the IPR, we only chose those atoms which contribute 10\% and more for a particular state labeled ($a\textendash f$). Those structures in the alloy which correspond to the selected mid-gap and band-tail states labeled ($a\textendash f$) are shown in Fig.~\ref{fig6}. As we can see from Fig.~\ref{fig4}, there is one mid-gap state and about five band-tail states. We estimated the mobility band-gap in this case to be ($\sim$1.6 eV).

The structures which are responsible for the mid-gap state labeled $c$, come from a threefold Si, a fivefold Ge, and geometrical defects. We have also obtained geometrical defects corresponding to three of the band-tail states labeled $(a,d, \text{and}~e)$. The larger IPR comes from one of these geometrical defects which contains 3 fourfold Si atoms connected to each others which have angular distribution off from a tetrahedral angle by $\pm$15$^\circ$. The larger contributions to the other two mid-gap states come from a fivefold Si atom together with a geometrical defect $(b)$, and a threefold Si bonded with a fivefold Si and a geometrical defect $(f)$.

In the case of $x$=0.4, the inverse participation ratio as a function of eigenvalue is plotted in Fig.~\ref{fig7}, while the contributions of each of the atoms to the IPR (only those atoms which contribute 5\% and more) for a particular state labeled ($a\textendash f$) and the different structures associated with these states are plotted in Fig.~\ref{fig8} and Fig.~\ref{fig9} respectively. As we can see from the IPR plots in the two cases, $x$=0.1 and $x$=0.4, as the Ge content increases, we observed an increasing number of band-tail states close to the conduction band edge and hence a narrow bandgap spectrum.

\begin{figure}
\begin{center}
\includegraphics[angle=0,width=0.48\textwidth]{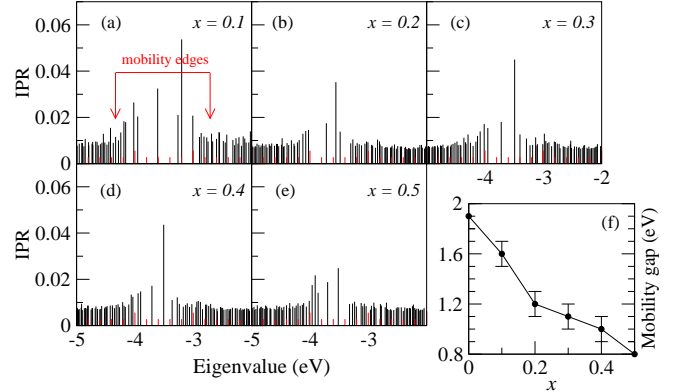}
\caption{\label{fig10} Inverse participation ratio (IPR) versus the energy eigenvalues for different $x$: (a)-(e) and the estimated mobility gap plotted versus the Ge concentration $x$: (f), in the a-Si$_{1-x}$Ge$_x$:H alloy.} 
\end{center}
\end{figure}

Further analysis of the type of bonds and structures associated with the mid-gap states and band-tail states showed that these states are coming not only from the dangling bonds defects (threefold and twofold bonds) but also from geometrical defects where there exist a strain bond angle in the structure which has an angular distribution far ($\pm 15^\circ$) from a tetrahedral angle. As compared with the case where $x$=0.1, we have observed an increase in the contribution from the geometrical defects to band-tail states for $x$=0.4. As shown in Fig.~\ref{fig9}, the larger contributions to the state labeled $(a~\text{and}~b)$ come from geometrical defects, a fivefold Si, and a threefold Ge structures. States labeled $d~\text{and}~f$ in this case are dominated by a geometrical defect which has 3 fourfold Si atoms connected to each other with strain bond with angular distribution off from a tetrahedral angle by $\pm$20$^\circ$. The number of Ge atoms in these geometrically defected structures has increased as compared with the case $x$=0.1. The dominant contributions to the mid-gap state arise from a fivefold Si atom bonded with two geometrical defects.

To emphasize the effect of Ge concentration in the mobility band gap, we have plotted the mobility gap as a function of the Ge concentration $x$ in Fig.~\ref{fig10} (f). The mobility gaps are extracted from Fig.~\ref{fig10} (a)-(e) which show the inverse participation ratio as a function of eigenvalues for different $x$. We observed a decrease in the mobility gap as the Ge concentration increase from $x$=0.1 to $x$=0.5. 
 The mobility gap is estimated by examination of the plots of IPR versus energy. In each case, there is a reasonable well-defined energy near the valence and conduction tails at which the IPR becomes roughly constant reflecting extended states. We include ``error bars" to convey a rough estimate of uncertainty in our estimated gaps.

\section{Dynamical properties}
\label{secV}

The dynamical properties of a-Si$_{1-x}$Ge$_x$:H alloys are analyzed with the vibrational density of states (VDOS) and inverse participation ratio. The vibrational  eigenvalues and eigenvectors are obtained from the dynamical matrix. The dynamical matrix is determined by displacing each atom with 0.03~\AA~in three orthogonal directions and then performing first principles force calculations for all the atoms for such displacement. Each calculation yields a column of force constant matrix. The vibrational eigenvectors and eigenvalues of the supercell are then obtained by diagonalizing the dynamical matrix. 

\begin{figure}[h]
\begin{center}
\includegraphics[angle=0,width=0.5\textwidth]{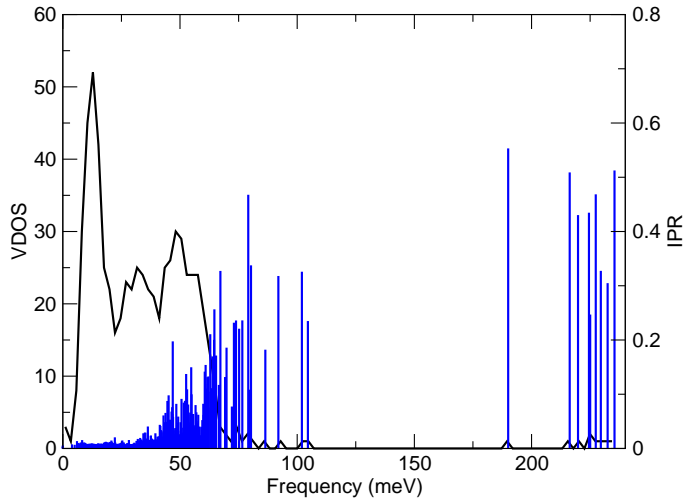}
\caption{\label{fig11}(Color Online) Vibrational density of states (black) and the inverse participation ratio (blue) for the a-Si$_{1-x}$Ge$_x$:H model for $x$=0.4. } 
\end{center}
\end{figure}

In Fig.~\ref{fig11}, the phonon density of state for a-Si$_{1-x}$Ge$_x$:H for $x$=0.4 is plotted. The acoustic peak appears at 13.29 meV and two optical peaks appear at  31.98 meV and 49.01 meV which is in good agreement with the experimental result reported by Mackenzie {\it et al} \cite{mackenzie}. The higher frequency modes in the range (213 meV-236 meV) are associated with hydrogen atoms with H\textendash Si and H\textendash Ge bonds which is in agreement with the experimental result reported by Wells {\it et al.} \cite{wells}.  The principal hydrogen related features of the spectrum which exhibit higher IPR (highly localized states) are: stretch modes of Si\textendash H at 2039 cm$^{-1}$, 2013 cm$^{-1}$, 1980 cm$^{-1}$, and 1873 cm$^{-1}$, and of Ge\textendash H at 1904 cm$^{-1}$, 1646 cm$^{-1}$ and a wagging vibration modes of Si\textendash H and Ge\textendash H dominates in the region of 600 \textendash~900 cm$^{-1}$.

\section*{CONCLUSION}

We have presented an {\it ab initio} first principles molecular dynamics study of a-Si$_{1-x}$Ge$_x$:H alloys for five different Ge atomic compositions. In this study, we reported a systematic investigations of structural and electronic properties of a-Si$_{1-x}$Ge$_x$:H alloys and the consequences of Ge atomic composition in the structure and electronic properties of the alloy. Where the structural properties of a-Si$_{1-x}$Ge$_x$:H alloys are concerned our results definitively show (a) very small change in the total coordination and total bond angle distribution, and (b) the emergence of preferential structures in the network as the Ge atomic content increases. The electronic density of states shows an increase in the band-tail states and narrow band gap as the Ge content increases due to the increase in the number of defects. This suggest the potential of Ge (by varying its atomic compositions) in tuning the band gap of the a-Si$_{1-x}$Ge$_x$:H alloy for different technological applications.    

\section{Acknowledgements}
We acknowledge the support from the Army Research Office and the National Science Foundation.

\end{document}